\documentclass{emulateapj}
\usepackage{apjfonts}
\usepackage{rotating}
\usepackage{enumerate}

\newcommand{\pivec}{\mbox{\boldmath $\pi$}}

\renewcommand{\thefootnote}{\ifcase\value{footnote}\or{$\dagger$} \or{$\ddagger$} \or($\infty$)\fi}

\lefthead{} 
\righthead{}

\begin{document}
\title{
OGLE-2016-BLG-0168 Binary Microlensing Event: Prediction and Confirmation of 
the Micorlens Parallax Effect from Space-based Observation
}

% Authort List ------------------------------------------------------------------------------------------------------------
\author{
% Leading Authors -------------------------------------------------------------
I.-G.~Shin\altaffilmark{1,17},
A.~Udalski\altaffilmark{2,16},
J.~C.~Yee\altaffilmark{1,17,18},
S.~Calchi~Novati\altaffilmark{8,9,18},
C.~Han\altaffilmark{15}\\
AND\\
% OGLE ------------------------------------------------------------------------
J.~Skowron\altaffilmark{2},
P.~Mr\'oz\altaffilmark{2},
I.~Soszy\'nski\altaffilmark{2},
R.~Poleski\altaffilmark{2,13},
M.~K.~Szyma\'nski\altaffilmark{2},
S.~Koz{\l}owski\altaffilmark{2},
P.~Pietrukowicz\altaffilmark{2},
K.~Ulaczyk\altaffilmark{2,3},
M.~Pawlak\altaffilmark{2}\\
(OGLE Collaboration),\\
% KMTNet ----------------------------------------------------------------------
M.~D.~Albrow\altaffilmark{6},   % Science part 
A.~Gould\altaffilmark{4,13,14}, 
S.-J.~Chung\altaffilmark{4,5},
K.-H.~Hwang\altaffilmark{4},
Y.~K.~Jung\altaffilmark{1},
Y.-H.~Ryu\altaffilmark{4},
W.~Zhu\altaffilmark{13},
S.-M.~Cha\altaffilmark{4,7},   % Tech part
D.-J.~Kim\altaffilmark{4},
H.-W.~Kim\altaffilmark{4,5},
S.-L.~Kim\altaffilmark{4,5},
C.-U.~Lee\altaffilmark{4,5},
Y.~Lee\altaffilmark{4,7},
B.-G.~Park\altaffilmark{4,5},
R.~W.~Pogge\altaffilmark{13}\\
(KMTNet Group),\\
% Spitzer ---------------------------------------------------------------------
C.~Beichman\altaffilmark{10},
G.~Bryden\altaffilmark{11},
S.~Carey\altaffilmark{12},
B.~S.~Gaudi\altaffilmark{13},
C.~B.~Henderson\altaffilmark{11,19},
Y.~Shvartzvald\altaffilmark{11,19}\\
({\it Spitzer} Team)\\
% -----------------------------------------------------------------------------
}

\bigskip\bigskip
% -------------------------------------------------------------------------------------------------------------------------
\affil{$^{1}$Smithsonian Astrophysical Observatory, 60 Garden St., Cambridge, MA 02138, USA}

% OGLE ------
\affil{$^{2}$Warsaw University Observatory, Al. Ujazdowskie 4, 00-478 Warszawa, Poland}
\affil{$^{3}$Department of Physics, University of Warwick, Gibbet Hill Road, Coventry CV4 7AL, UK}

% KMTNet ------
\affil{$^{4}$Korea Astronomy and Space Science Institute, 776 Daedeokdae-ro, Yuseong-Gu, Daejeon 34055, Korea}
\affil{$^{5}$Korea University of Science and Technology, 217 Gajeong-ro, Yuseong-gu, Daejeon 34113, Korea}
\affil{$^{6}$University of Canterbury, Department of Physics and Astronomy, Private Bag 4800, Christchurch 8020, New Zealand}
\affil{$^{7}$School of Space Research, Kyung Hee University, Giheung-gu, Yongin, Gyeonggi-do, 17104, Korea}

% Spitzer -------
\affil{$^{8}$ IPAC, Mail Code 100-22, California Institute of Technology, 1200 E. California Boulevard, Pasadena, CA 91125, USA}
\affil{$^{9}$ Dipartimento di Fisica ``E. R. Caianiello'', Universit\a di Salerno, Via Giovanni Paolo II, I-84084 Fisciano (SA), Italy}
\affil{$^{10}$ NASA Exoplanet Science Institute, California Institute of Technology, Pasadena, CA 91125, USA}
\affil{$^{11}$ Jet Propulsion Laboratory, California Institute of Technology, 4800 Oak Grove Drive, Pasadena, CA 91109, USA}
\affil{$^{12}$ Spitzer Science Center, MS 220-6, California Institute of Technology, Pasadena, CA, USA}

% Etc ----
\affil{$^{13}$Department of Astronomy, Ohio State University, 140 W. 18th Ave., Columbus, OH 43210, USA}
\affil{$^{14}$Max-Planck-Institute for Astronomy, K\"onigstuhl 17, 69117 Heidelberg, Germany}
\affil{$^{15}$Department of Physics, Chungbuk National University, Cheongju 371-763, Republic of Korea}

% -------------------------------------------------------------------------------------------------------------------------
\footnotetext[16]{OGLE Collaboration}
\footnotetext[17]{KMTNet Group}
\footnotetext[18]{{\it Spitzer} Team}
\footnotetext[19]{NASA Postdoctoral Program Fellow}

% 0. ABSTRACT -----------------------------------------------------------------------------------------
\begin{abstract}
 The microlens parallax is a crucial observable for conclusively identifying the nature of lens systems 
in microlensing events containing or composed of faint (even dark) astronomical objects such as planets, 
neutron stars, brown dwarfs, and black holes. With the commencement of a new era of microlensing in 
collaboration with space-based observations, the microlens parallax can be routinely measured. In 
addition, space-based observations can provide opportunities to verify the microlens parallax measured 
from ground-only observations and to find a unique solution of the lensing lightcurve analysis. However, 
since most space-based observations cannot cover the full lightcurves of lensing events, it is also 
necessary to verify the reliability of the information extracted from fragmentary space-based lightcurves. 
We conduct a test based on the microlensing event OGLE-2016-BLG-0168 created by a binary lens system 
consisting of almost equal mass M-dwarf stars to demonstrate that it is possible to verify the microlens 
parallax and to resolve degeneracies by using the space-based lightcurve even though the observations are 
fragmentary. Since space-based observatories will frequently produce fragmentary lightcurves due to their 
short observing windows, the methodology of this test will be useful for next-generation microlensing 
experiments that combine space-based and ground-based collaboration.
\end{abstract}


% 1. INTRODUCTION -------------------------------------------------------------------------------------
\section{Introduction}

 The microlensing technique can probe a variety of astronomical objects in a wide range of masses such 
as planets, neutron stars, brown dwarfs, and isolated black holes \citep{dong07, miyake12, poindexter05, 
shvartzvald15, wyrzykowski16}. The microlensing technique can detect these faint or dark objects 
regardless of their luminosity levels, in sharp contrast to other methods, which as a matter of course 
are restricted to studying objects within their flux detection limits. 
 
 To conclusively reveal the nature of the lens system that generates a microlensing event, additional 
observables are required such as the microlens parallax, $\pi_{\rm E}$, and the angular Einstein ring 
radius, $\theta_{\rm E}$. Based on these additional observables, the properties of the lens system can 
be determined from
% Equation (1) ----------------------------------------------------------
\begin{equation}
M_{\rm L} = {\theta_{\rm E} \over \kappa \pi_{\rm E}} ~~;~~ 
D_{\rm L}= { {\rm AU} \over {\pi_{\rm E} \theta_{\rm E} + \pi_{\rm S} } },
\end{equation}
% -----------------------------------------------------------------------
where $M_{\rm L}$ is the total mass of the lens system, $D_{\rm L}$ is the distance to the lens system 
toward the Galactic bulge, $\pi_{\rm S}$ is the parallax of the background star (source) defined as 
$\pi_{\rm S}={\rm AU}/D_{\rm S}$ where the $D_{\rm S}$ is the distance to the source, and $\kappa \equiv 
4G/(c^2{\rm AU}) \sim 8.1~{\rm mas}/M_{\odot}$. Although $\pi_{\rm E}$ and $\theta_{\rm E}$ appear 
equally important in Equation~(1), $\pi_{\rm E}$ is actually more crucial because $\theta_{\rm E}$ is 
easily determined from the finite source effect with high-cadence observations. In particular, for a 
binary lensing event, $\theta_{\rm E}$ can be routinely measured when the source crosses or approaches 
caustics of binary lensing events. Thus, it is important to securely and accurately measure the microlens 
parallax. 

 However, the measurement of the microlens parallax based on ground-only observations is made from subtle 
deviations in those lensing lightcurves that have a sufficiently long time-scale to make manifest the 
deviations caused by Earth's orbital motion. As a result, there exist some obstacles to measuring the 
microlens parallax. First, the signal of the microlens parallax, i.e., subtle deviations in the lightcurve, 
can be detected if Earth moves enough to produce the signal over the duration of the event. Thus, the 
microlens parallax can be measured for only some cases of lensing events that have long time-scales 
(usually, $t_{\rm E}\geq30$ days). Second, the measurement can be confused with systematics that can make 
a false positive detection or inaccurate measurement of the microlens parallax. Third, there exist 
degeneracies in the microlens parallax that prevent accurately or uniquely measuring it. For example, 
the ecliptic degeneracy \citep{jiang04,skowron11} produces degenerate solutions with different values of 
the microlens parallax that can describe the same lensing lightcurve. Also, the lens-orbital effect 
caused by orbital motion of the lens components affects the measured values of the microlens parallax 
\citep{batista11,shin12,skowron11}. Hence, before the era of space-based microlensing, the microlens 
parallax could be securely and accurately measured for only a small number of lensing events that satisfy 
conditions to measure it during a bulge season.

 In the new era, however, the microlens parallax can be routinely and securely measured in collaboration 
with space-based observations. In principle, the offset between ground and space telescopes provides a 
chance to routinely measure the microlens parallax regardless of the magnification level of the lensing 
event. In addition, space-based observations can provide opportunities to verify the measurement of the 
microlens parallax and to resolve degeneracies in the microlens parallax.

 However, for lensing events having a relatively long time-scale, space-based observations can cover 
only fragmentary parts of the full lensing lightcurve due to short observing windows. For example, the 
{\it Spitzer} space telescope has only a $\sim40$ day observing window. Moreover, space-based 
observations generally do not cover caustic-crossing features of the binary lensing event because it is 
almost impossible to predict the exact time when the source crosses the caustic structure, especially 
for long-time scale events. Indeed, for single lensing events, \citet{yee15b} posit and 
\citet{calchi15a} and \citet{zhu17} show that fragmentary lightcurves can be successfully exploited to 
extract microlens parallaxes for point-lens (or near point-lens) events. However, this has not been 
demonstrated for binary lensing events.

Because these fragmentary space-based lensing lightcurves are quite common, it is important to do a test 
whether it is possible to extract reliable information from them or not. In fact, during the {\it Spitzer} 
microlensing campaign in $2015-2016$, most of the observed lightcurves are fragmentary. Thus, we conduct 
such a test by using the binary microlensing event OGLE-2016-BLG-0168 which has {\it Spitzer} observations. 
The event has a long time-scale ($t_{\rm E}\sim 90$ days) and the {\it Spitzer} observations covered a short 
part ($\sim 30$ days) of the full lensing lightcurve. Moreover, we found degenerate solutions to the event 
during the analysis. As a result, this event is a perfect test bed to show the possibility of extracting 
information from the fragmentary lightcurve observed by a space-based observatory. Our test can provide an 
important example to probe the reliability of extracting information from the fragmentary space-based 
observations. In addition, the methodology of this test can provide procedures to systematically measure 
and verify the microlens parallax based on fragmentary lightcurves from space and to resolve the degenerate 
solutions, especially for the {\it Spitzer} microlensing campaign. 

In this paper, we describe observations of the event in Section 2. In Section 3, we describe our analysis 
procedures and the test. In Section 4, we present results of the analysis and the test of the event. 
Lastly, we discuss and summarize the results in Section 5.

% 2. OBSERVATIONS -------------------------------------------------------------------------------------
\section{Observations}

\begin{figure}[ht]
\epsscale{1.10}
\plotone{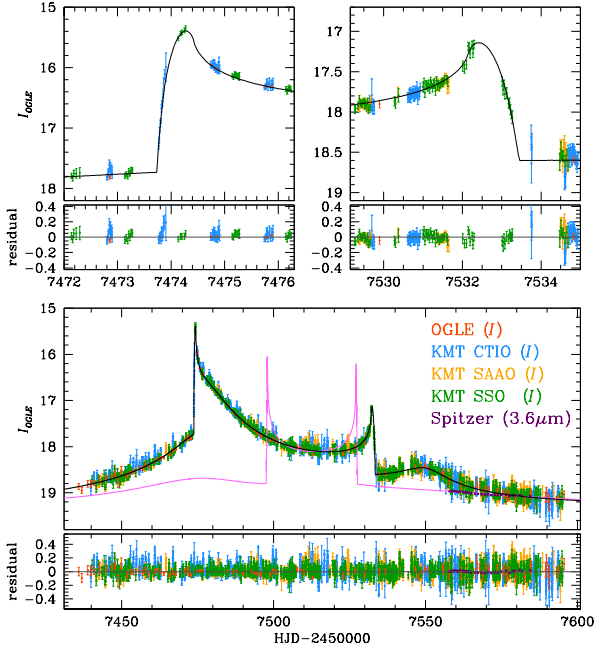}
\caption{\label{fig:one}
Lightcurves of the binary microlensing event OGLE-2016-BLG-0168. Each color represents observed data 
from different telescopes located in ground and space. Black and pink solid lines indicate the model 
lightcurves of ground and {\it Spitzer} observations, respectively. Upper panels show the zoom-ins 
of the ground lightcurve the caustic entrance (left) and exit (right). Lower panels show the whole 
lightcurves with residuals between models and observations.
}\end{figure}

The microlensing event OGLE-2016-BLG-0168 occurred on the source star located in the galactic bulge at 
$(\alpha,\delta)_{\rm J2000}= (17^{h}50^{m}49^{s}.89,-31^{\circ}45^{'}30^{''}.1)$ in equatorial 
coordinates and $(l,b)=(-1.84,-2.42)$ in galactic coordinates. The event was observed 
both by ground-based surveys and the {\it Spitzer} space telescope. In Figure \ref{fig:one}, we present 
the observed lightcurve of OGLE-2016-BLG-0168. The upper two panels show the caustic-crossing parts of 
the lightcurve and the lower panels show the entire duration of significant magnification. The lightcurve 
observed from ground-based telescopes shows typical features caused by a binary lens system.

% 2.1. OBSERVATIONS: GROUND-BASED OBSERVATIONS --------------------------------------------------------
\subsection{Ground-based observations}

 The event was announced by the Optical Gravitational Lensing Experiment \citep[OGLE:][]{udalski15a} based 
on observations with its 1.3 m Warsaw telescope with $1.4~{\rm deg^2}$ camera located at the Las Campanas 
Observatory in Chile. The event was alerted by the Early Warning System \citep[EWS:][]{udalski94,udalski03} 
of the OGLE survey on 2016 February 21. The OGLE data in {\it I-}band were reduced by a pipeline based on 
the Difference-Imaging Analysis method \citep{alard98, wozniak00}. The uncertainties of the OGLE data were 
re-scaled according to the description in \citet{skowron16}. 

 The Korea Microlensing Telescope Network \citep[KMTNet:][]{kim16} survey, which is designed for high-cadence 
monitoring toward the galactic bulge with large a field-of-view, independently observed the event. KMTNet is 
a telescope network consisting of three identical 1.6 m telescopes with $4~{\rm deg^2}$ cameras located at 
Cerro Tololo Inter-American Observatory in Chile (KMTC), South African Astronomical Observatory in South 
Africa (KMTS), and Siding Spring Observatory in Australia (KMTA). For the event, KMTC and KMTA observations 
cover the caustic entrance (${\rm HJD-2450000=HJD'} \sim 7474.2$) and exit (${\rm HJD'} \sim 7532.5$) parts 
of the lightcurve with $\sim15$ minute cadence. KMTNet data in {\it I-}band were reduced by pySIS 
\citep{albrow09}, which employs the image subtraction method. 

The event was also observed in both {\it I-} and {\it H-}band by the SMARTS 1.3 m telescope at CTIO in Chile. 
These data were not used in the modeling, but were used to determine the $({\it I-H})$ source color 
(see Section 4.4).

% Figure 1 ----------------------------------------------------------------------------------------
% --------------------------------------------------------------------------------------------------

% 2.2. OBSERVATIONS: SPACE-BASED OBSERVATIONS ---------------------------------------------------------
\subsection{Space-based observations}

 The event was observed by the {\it Spitzer} space telescope with the $3.6~\mu m$ channel (hereafter, 
{\it L-}band) of the IRAC camera. Briefly, the event was selected on 2016 June 16 as a subjective 
target based on the selection criteria described in \citet{yee15b} because the lightcurve from 
ground-based observations showed typical anomaly features caused by the binary lens system. 
The observations started on 2016 June 18 (${\rm HJD'} \sim 7557.93$) and ended July 14 (${\rm HJD'} 
\sim 7584.48$). During $4$ weeks of observations with cadence $\sim 1~{\rm day^{-1}}$, $28$ data 
points of the event were gathered and then the data were reduced by using methods described in 
\citet{calchi15b}.

% Figure 2 ----------------------------------------------------------------------------------------
\begin{figure}[ht]
\epsscale{1.10}
\plotone{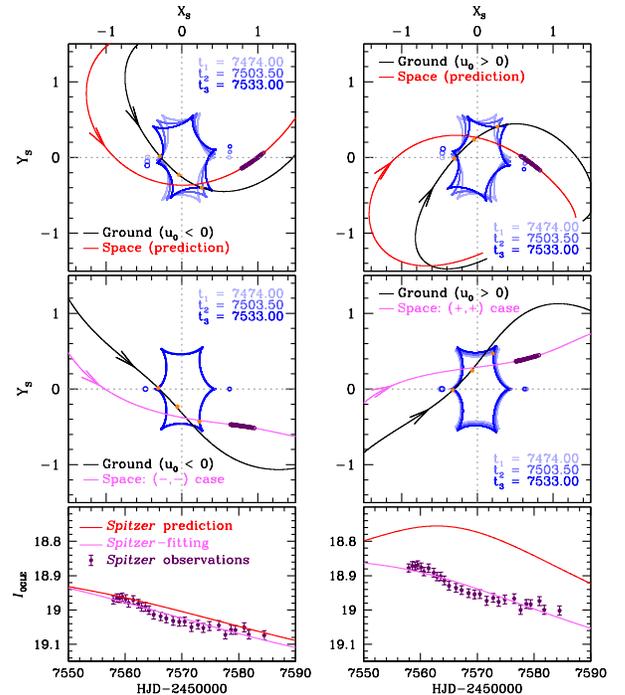}
\caption{\label{fig:two}
Geometries of the binary microlensing event OGLE-2016-BLG-0168. Top panels show caustic features 
reflecting the orbital motion of the binary lens system at the time when the source enters $({\rm 
HJD'}\sim7474.0)$, is inside $(\sim7503.5)$, and exits $(\sim7533.0)$ the caustic; the orange dots 
mark the source position at those times. In the panels, the black line with an arrow indicates the 
source trajectory of ground-based models and the red line indicates the predicted source trajectory 
of the {\it Spitzer} lightcurve based on the ground models of $(u_0<0)$ (left) and $(u_0>0)$ (right) 
cases. The purple points represent the coverage of {\it Spitzer} observations. Middle panels show 
geometries of models with combined data from ground and {\it Spitzer}. The color scheme is the same as 
the top panel except that now a pink line indicates the {\it Spitzer}-fitted lightcurve. Bottom panels 
show the prediction, {\it Spitzer}-fitting, and observations of {\it Spitzer} lightcurve for both 
$(-,-)$ (left) and $(+,+)$ (right) cases, respectively. 
}\end{figure}
% --------------------------------------------------------------------------------------------------

% 2.3. OBSERVATIONS: EXTINCTION -----------------------------------------------------------------------
\subsection{Extinction}

 The source star of the event is located in a severely extincted field. The source extinction is 
$A_I \sim 4.9$ in {\it I-}band and $A_L \sim 0.35$ in {\it L-}band \footnote[1]{The $A_I$ value is 
measured from the CMD analysis of this event (see Section 4.4). Based on the {\it I-}band extinction, 
the $A_L$ value is calculated by using the relationship between optical and infrared extinction 
\citep{cardelli89}.}. As a result, the source is relatively faint for ground-based observations from 
OGLE and KMTNet. In contrast to ground-based observations, the source is a quite bright target for 
{\it Spitzer} observations.

%$A_{3.6~\mu m}/A_{K}=(3.6/2.2)^{-1.61},~A_{\it K}=0.15A_{I}$ 

% 3. ANALYSIS -----------------------------------------------------------------------------------------
\section{Analysis}

 We model the lightcurves of the OGLE-2016-BLG-0168 event to reveal the nature of the binary lens 
system causing the microlensing event. In addition, we conduct a test to validate the microlens 
parallax and resolve the degeneracy in the microlens parallax.

% Table 1 ------------------------------------------------------------------
\begin{deluxetable*}{lrrrrr}
\tablecaption{The best-fit models of the ground-based observations\label{table:one}}
\tablewidth{0pt}
\tablehead{
% ---------------------------------------------------------------------------
\multicolumn{1}{c}{} &
\multicolumn{1}{c}{} &
\multicolumn{2}{c}{$(u_0<0)$} &
\multicolumn{2}{c}{$(u_0>0)$} \\
\multicolumn{1}{c}{parameter} &
\multicolumn{1}{c}{STD} &
\multicolumn{1}{c}{PRX} &
\multicolumn{1}{c}{OBT+PRX} &
\multicolumn{1}{c}{PRX} &
\multicolumn{1}{c}{OBT+PRX} 
}
\startdata
% -------------------------------------------------------------------------------------------------------------------------------------------------
$\chi^2 / {\rm dof}$       & 6501.53 / $(6232-7)$ & 6342.13 / $(6232-9)$ & 6227.09 / $(6232-11)$ & 6348.94 / $(6232-9)$ & 6240.25 / $(6232-11)$ \\
$t_0$ (HJD')               & 7492.261 $\pm$ 0.144 & 7492.636 $\pm$ 0.417 & 7492.478 $\pm$ 0.547  & 7492.188 $\pm$ 0.414 & 7492.595 $\pm$ 0.420  \\
$u_0$                      &    0.199 $\pm$ 0.002 &   -0.202 $\pm$ 0.003 &   -0.201 $\pm$ 0.005  &    0.199 $\pm$ 0.003 &    0.207 $\pm$ 0.004  \\
$t_{\rm E}$ (days)         &   89.786 $\pm$ 0.232 &   88.525 $\pm$ 0.316 &   97.010 $\pm$ 1.345  &   88.550 $\pm$ 0.308 &   95.379 $\pm$ 1.024  \\
$s$                        &    1.120 $\pm$ 0.001 &    1.117 $\pm$ 0.001 &    1.075 $\pm$ 0.014  &    1.115 $\pm$ 0.001 &    1.092 $\pm$ 0.008  \\
$q$                        &    0.632 $\pm$ 0.009 &    0.664 $\pm$ 0.021 &    0.724 $\pm$ 0.024  &    0.648 $\pm$ 0.020 &    0.736 $\pm$ 0.024  \\
$\alpha$ (rad)             &    5.448 $\pm$ 0.002 &   -5.462 $\pm$ 0.006 &   -5.379 $\pm$ 0.009  &    5.454 $\pm$ 0.006 &    5.367 $\pm$ 0.007  \\
$\rho_{\star}$ ($10^{-2}$) &    0.372 $\pm$ 0.007 &    0.375 $\pm$ 0.006 &    0.400 $\pm$ 0.007  &    0.378 $\pm$ 0.007 &    0.395 $\pm$ 0.007  \\
$\pi_{{\rm E},{\it N}}$    &          ---         &    0.033 $\pm$ 0.004 &    0.382 $\pm$ 0.022  &   -0.038 $\pm$ 0.003 &   -0.475 $\pm$ 0.025  \\
$\pi_{{\rm E},{\it E}}$    &          ---         &    0.013 $\pm$ 0.009 &    0.057 $\pm$ 0.011  &    0.014 $\pm$ 0.009 &   -0.026 $\pm$ 0.011  \\
$ds/dt$ ($yr^{-1}$)        &          ---         &          ---         &    0.287 $\pm$ 0.120  &          ---         &    0.090 $\pm$ 0.067  \\
$d\alpha/dt$ (rad/yr)      &          ---         &          ---         &   -1.437 $\pm$ 0.138  &          ---         &    1.574 $\pm$ 0.121  \\
$F_{\rm S, OGLE}$          &    0.248 $\pm$ 0.001 &    0.251 $\pm$ 0.001 &    0.252 $\pm$ 0.001  &    0.250 $\pm$ 0.001 &    0.253 $\pm$ 0.001  \\
$F_{\rm B, OGLE}$          &    0.048 $\pm$ 0.001 &    0.045 $\pm$ 0.001 &    0.044 $\pm$ 0.001  &    0.046 $\pm$ 0.001 &    0.043 $\pm$ 0.001  \\ 
% --------------------------------------------------------------------------------------------------------------------------------------------------
\enddata
\tablecomments{
${\rm HJD' = HJD-2450000}$,
Abbreviations -- 
STD: the static model, 
PRX: the model considering the annual microlens parallax,
OBT: the model considering the lens-orbital motion.
}
\end{deluxetable*}
% ---------------------------------------------------------------------------
 
 Because the event was simultaneously observed by ground and space telescopes, we try to find fits 
for both observed lightcurves by using parameters adopted from the conventional parameterization 
\citep{refsdal66, gould92, gould94, graff02, shin13, jung15, udalski15b, zhu15}. We briefly summarize 
the parameterization to facilitate further description of the modeling. We used in total $11$ geometric 
parameters to construct model lightcurves considering the higher-order effects. Among these, $7$ parameters 
($t_0$, $u_0$, $t_{\rm E}$, $s$, $q$, $\alpha$, and $\rho_{\ast}$) are used to describe the static binary 
lens model. The other $4$ parameters are used to describe vector $\pivec_{\rm E}$ components ($\pi_{{\rm E}, 
{\it N}}$, $\pi_{{\rm E}, {\it E}}$) of the microlens parallax and orbital motion ($ds/dt$, $d\alpha/dt$) 
of the binary lens components. For parameters of the static binary lens model, $t_0$, $u_0$, $t_{\rm E}$, 
and $\alpha$ are related to describing of the trajectory of the magnified background star (hereafter, 
source) as seen from the ground, which are defined as the time of the closest source approach to the 
center of mass of the binary lens system, the impact parameter (separation between the center of mass 
and the source position at time of $t_0$), the source crossing time along the angular Einstein ring 
radius, i.e., $\theta_{\rm E}$, and the angle of the source trajectory with respect to the binary axis, 
respectively. The parameters $s$ and $q$ are related to describing the caustic structure and are defined as 
the projected separation between the binary stars normalized by $\theta_{\rm E}$, and the mass ratio of 
the primary and secondary stars, respectively. The last parameter $\rho_{\ast}$ is defined as the source 
radius normalized by $\theta_{\rm E}$, i.e., $\rho_{\ast}=\theta_{\ast}/\theta_{\rm E}$, which can 
provide a measurement of $\theta_{\rm E}$ based on the finite source effect that moderates the amplitude 
of magnification when the source crosses the caustics.

 The modeling sequence consists of three phases. In the first phase, to find a global minimum, we conduct 
a grid search of the $(s,q)$ parameter space because the parameters are directly related to the caustic 
structure, which leads to dramatic changes in features of the static binary model lightcurve. For the other 
$5$ basic parameters, we allow that these parameters can be varied from proper initial values to fit the 
observed lightcurve by using the $\chi^2$ minimization method called Markov Chain Monte Carlo (MCMC) algorithm. 
In the second phase, based on the static binary model found in the first phase, we sequentially introduce 
the higher-order effects caused by the microlens parallax and the orbital motion of the binary lens components. 
These effects can produce better fits if there exist residuals between the static model and the observed 
lightcurve. Note that both effects should be simultaneously considered because both simultaneously affect 
the curvature of the source trajectory and reflect physical motions. In the last phase, we refine the models 
after re-scaling the errors of the observed data based on the best-fit model, so that each data point can be 
represented as $\sim 1~ \chi^2 / {\rm dof}$ when the models are computed. During the refining process, we 
consider the variation of the magnification due to the limb-darkening of the source's surface by adopting 
coefficients from \citet{claret00} that correspond to the source type of the event (in Section 4.4, 
determining the source type is described in detail). In this phase, we allow all parameters to vary in wide 
ranges to estimate their uncertainty based on scatter of the MCMC chain.

% 3.1. ANALYSIS: MODELING OF THE GROUND-BASED LIGHTCURVE ----------------------------------------------
\subsection{Modeling of the ground-based lightcurve}

 In Figure \ref{fig:one}, we present observed lightcurves as seen from ground and space. The lightcurve 
shows a typical ``U''-shape of a binary lensing lightcurve. As shown in the zoom-ins, the caustic entrance 
and exit are well-covered by the KMTNet survey and thus we can clearly measure the angular Einstein ring 
radius. In addition, the time between the caustic entrance and exit is $\sim 60$ days. This is long enough 
to expect to detect signals in the ground-based lightcurve caused by the annual microlens parallax and 
lens-orbital effects. In Figure \ref{fig:two}, we also present the geometry of the event, which is described 
in detail in Sections 3.2 and 4.2.

 In Table \ref{table:one}, we present models with best-fit parameters of the degenerate solutions considering 
the lens-parallax and lens-orbital effects. We find that there exist two degenerate models, $(u_0<0)$ and 
$(u_0>0)$, for the ground-based lightcurve. In the best-fit models, signals of the microlens parallax and 
the lens-orbital effects are clearly detected. When the microlens parallax effect \citep[annual microlens 
parallax effect:][]{gould92} is introduced, we find that the $\chi^2$ improvements compared to the static 
model are $159.4$ and $152.6$ for the $(u_0<0)$ and $(u_0>0)$ cases, respectively. In addition, when we 
supplement the lens-parallax model with the lens-orbital effect \citep[approximated lens-orbital effect:][]{shin13}, 
we find that the $\chi^2$ improvements are $115.0$ and $108.7$ for the $(u_0<0)$ and $(u_0>0)$ cases, respectively. 
It implies that the lens-orbital effect is clearly detected for both cases. Note that the signal of the lens-orbital 
effect comes from the ground-based observations. This signal is quite strong because the orbital motion of the lens 
components changes the caustic structure and thus the signal comes from the caustic parts which are covered by ground 
surveys, especially the KMTNet survey. Note that, since clear signal of the lens-orbital effect is detected, we 
investigated complete Keplerian orbital solutions \citep[parameters adopted from][]{shin11,skowron11}. However, 
the Kepler parameters could not be meaningfully constrained for this case. The $\chi^2$ difference between the 
best-fit models in Table \ref{table:one} is only $\chi^2(u_0>0)-\chi^2(u_0<0)=13.4$. We note that, for this event, 
there do not exist degenerate solutions caused by the close/wide degeneracy \citep{griest98, dominik99, an05} 
because the best-fit models have resonant caustics ($s \sim 1$).

% Table 2 -------------------------------------------------------------------
\begin{deluxetable*}{lrrrr}
\tablecaption{The best-fit models of the combined observations\label{table:two}}
\tablewidth{0pt}
\tablehead{
% ---------------------------------------------------------------------------
\multicolumn{1}{c}{} &
\multicolumn{2}{c}{$(-,-)$} &
\multicolumn{2}{c}{$(+,+)$} \\
\multicolumn{1}{c}{parameter} &
\multicolumn{1}{c}{w/o cc} &
\multicolumn{1}{c}{w/ cc} &
\multicolumn{1}{c}{w/o cc} &
\multicolumn{1}{c}{w/ cc} 
}
\startdata
% ----------------------------------------------------------------------------------------------------------------------------------------------------
$\chi^2_{\rm total}   / {\rm dof}$      & 6257.46 / $(6260-11)$ & \bf{6259.74 / $\bf (6260-11)$      } & 6296.21 / $(6260-11)$ & 6296.16 / $(6260-11)$      \\
$\chi^2_{\rm Ground}  / {\rm N_{data}}$ & 6227.88 / 6232        & \bf{6230.48 / 6232                 } & 6257.18 / 6232        & 6256.03 / 6232             \\
$\chi^2_{\it Spitzer} / {\rm N_{data}}$ & 29.58 / 28            & \bf{  29.26 / 28                   } & 39.03 / 28            &   40.13 / 28               \\
$\chi^2_{\rm penalty}$                  & ---                   & \bf{2.98 ($\bf <2\sigma_{\rm cc}$) } & ---                   & 0.14 ($<1\sigma_{\rm cc}$) \\
$t_0$ (HJD')                            & 7492.470 $\pm$ 0.488  & \bf{7493.687 $\pm$ 0.311           } & 7493.130 $\pm$ 0.498  & 7493.298 $\pm$ 0.461       \\
$u_0$                                   &   -0.203 $\pm$ 0.005  & \bf{  -0.214 $\pm$ 0.003           } &    0.218 $\pm$ 0.005  &    0.220 $\pm$ 0.005       \\
$t_{\rm E}$ (days)                      &   95.249 $\pm$ 1.224  & \bf{  93.670 $\pm$ 1.165           } &   89.449 $\pm$ 1.036  &   89.395 $\pm$ 0.921       \\
$s$                                     &    1.088 $\pm$ 0.012  & \bf{   1.104 $\pm$ 0.011           } &    1.139 $\pm$ 0.011  &    1.140 $\pm$ 0.010       \\
$q$                                     &    0.713 $\pm$ 0.023  & \bf{   0.768 $\pm$ 0.015           } &    0.731 $\pm$ 0.027  &    0.741 $\pm$ 0.024       \\
$\alpha$ (rad)                          &   -5.389 $\pm$ 0.008  & \bf{  -5.403 $\pm$ 0.008           } &    5.401 $\pm$ 0.007  &    5.401 $\pm$ 0.005       \\
$\rho_{\star}$ ($10^{-2}$)              &    0.393 $\pm$ 0.007  & \bf{   0.396 $\pm$ 0.007           } &    0.388 $\pm$ 0.007  &    0.390 $\pm$ 0.007       \\
$\pi_{{\rm E},{\it N}}$                 &    0.349 $\pm$ 0.024  & \bf{   0.360 $\pm$ 0.027           } &   -0.401 $\pm$ 0.034  &   -0.411 $\pm$ 0.023       \\
$\pi_{{\rm E},{\it E}}$                 &    0.062 $\pm$ 0.011  & \bf{   0.047 $\pm$ 0.009           } &   -0.002 $\pm$ 0.010  &   -0.004 $\pm$ 0.010       \\
$ds/dt$ ($yr^{-1}$)                     &    0.182 $\pm$ 0.105  & \bf{   0.050 $\pm$ 0.094           } &   -0.325 $\pm$ 0.103  &   -0.337 $\pm$ 0.093       \\
$d\alpha/dt$ (rad/yr)                   &   -1.238 $\pm$ 0.123  & \bf{  -1.133 $\pm$ 0.125           } &    0.933 $\pm$ 0.103  &    0.954 $\pm$ 0.070       \\
$F_{\rm S, OGLE}$                       &    0.252 $\pm$ 0.001  & \bf{   0.254 $\pm$ 0.001           } &    0.253 $\pm$ 0.001  &    0.253 $\pm$ 0.001       \\
$F_{\rm B, OGLE}$                       &    0.044 $\pm$ 0.001  & \bf{   0.042 $\pm$ 0.001           } &    0.044 $\pm$ 0.001  &    0.043 $\pm$ 0.001       \\
$F_{\rm S, {\it Spitzer}}$              &   38.967 $\pm$ 1.059  & \bf{  36.357 $\pm$ 0.589           } &   28.773 $\pm$ 2.647  &   27.961 $\pm$ 1.692       \\
$F_{\rm B, {\it Spitzer}}$              &   -7.716 $\pm$ 1.519  & \bf{  -4.306 $\pm$ 0.845           } &    2.834 $\pm$ 3.339  &    3.726 $\pm$ 2.262       \\
% ----------------------------------------------------------------------------------------------------------------------------------------------------
\enddata
\tablecomments{
${\rm HJD' = HJD-2450000}$. See Section 3.3 for the definition of $\chi^2_{\rm penalty}$.
}
\end{deluxetable*}
% ---------------------------------------------------------------------------

% 3.2. ANALYSIS: THE MICROLENS PARALLAX TEST BASED ON THE SPITZER OBSERVATIONS ------------------------
\subsection{The microlens parallax test based on the {\it Spitzer} observations}

 Based on space-based observations, it is possible to conduct a test for verifying the 
measurement of the annual microlens parallax. In addition, as pointed out by \citet{han16a, han16b}, 
space-based observations can also provide an opportunity to resolve degenerate solutions. Thus, we 
conduct a test based on the {\it Spitzer} observations to verify the annual microlens parallax and 
lens-orbital motion effects from results of the ground-based models. In addition, we try to resolve 
degenerate solutions by using the {\it Spitzer} observations.

 In Figure \ref{fig:two}, the red lines indicate predicted source trajectories that should be seen by 
the {\it Spitzer} space telescope. These predicted source trajectories are produced by using ground-based 
models considering the annual microlens parallax and lens-orbital effects. By comparing the prediction 
without {\it Spitzer} and the fitting with {\it Spitzer} data, it is possible to check the measurement 
of the microlens parallax and lens-orbital motion. Note that the parameters of the annual and satellite 
microlens parallaxes are defined in the same reference frame \citep{calchi15a,yee15a,zhu15}. As a result, 
it is possible to directly compare the microlens parallax values. In addition, this validation process 
can provide a chance to resolve the degenerate solutions.

 As shown in Figure \ref{fig:two} (purple dots on the predicted trajectories), {\it Spitzer} observations 
covered only a short segment ($\sim 26$ days) compared to the total Einstein timescale of the event 
($\sim 94$ days). For relatively long timescale microlensing events, space-based observations usually 
cover only part of the lensing lightcurve due to the short observing window. As a result, this fragmentary 
{\it Spitzer} lightcurve is quite common for long time-scale lensing events. Thus, our test can provide 
an important example of whether it is possible to extract secure microlens parallax information from 
a fragmentary lightcurve or not.

% Figure 3 ----------------------------------------------------------------------------------------
\begin{figure*}[ht]
\epsscale{1.10}
\plotone{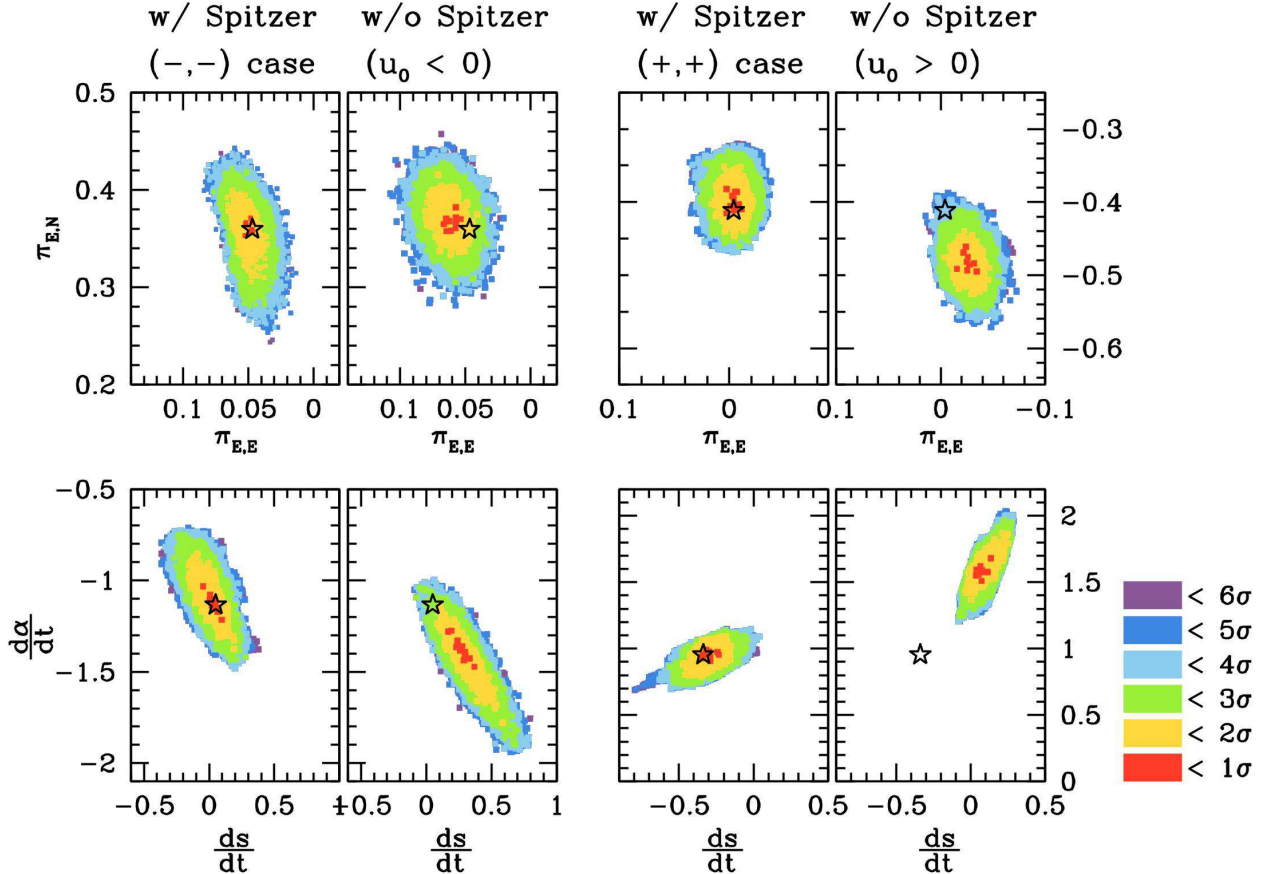}
\caption{\label{fig:three}
The distributions of the microlens parallax and lens orbital motion parameters. Left panels show 
$(-,-)$ and $(u_0<0)$ cases of the MCMC chain scatters with and without the {\it Spitzer} data. 
Red, yellow, green cyan, blue, and purple colors represents $\Delta \chi^2$ between the best-fit 
and chain value less than $1, 2, 3, 4, 5,$ and $6 \sigma$, respectively. The star represents 
the best-fit value of the model including the {\it Spitzer} data. Right panels show $(+,+)$ and 
$(u_0>0) $ cases with an identical scheme to the right panels. 
}\end{figure*}
% --------------------------------------------------------------------------------------------------

% 3.3. ANALYSIS: MODELING OF THE SPITZER LIGHTCURVE ---------------------------------------------------
\subsection{Modeling of {\it Spitzer} lightcurve}

 For this test, we conduct modeling of the combined ground and {\it Spitzer} observations. We present 
observed lightcurves with the best-fit models in Figure \ref{fig:one}. During the modeling process, we 
investigate degenerate solutions caused by the ``four-fold degeneracy'' \citep{zhu15}. The degeneracies 
are caused by different source trajectories seen by ground and space telescopes passing over a similar 
lensing magnification pattern, which is reflected over the binary axis. The degenerate solutions are 
denoted by the combination of the signs of impact parameters as seen from the space and ground, i.e., 
$(\pm, \pm)$, according to the conventional way \citep[see Section 3 in][]{zhu15}. Under our 
parameterization, we can control the source trajectories by changing the sign of $u_0$ and $\pi_{{\rm E},
{\it N}}$ parameters and thus we carefully set initial values to search for these models. For this event, 
we find that the $(-,-)$ and $(+,+)$ models showed similar fits with $\Delta\chi^2 \sim 36.4$. However, 
there do not exist plausible local minima of the $(+,-)$ and $(-,+)$ solutions \footnote[2]{For $(+,-)$ case, 
we found a plausible model but the $\chi^2$ of the model is larger than $\sim 150$ compared to the best-fit 
model. This $\Delta \chi^2$ is too large to claim the $(+,-)$ solution is a degenerate solution because 
there exist noticeable deviations between observed and model lightcurves. Thus, the $(+,-)$ model is 
rejected as one of degenerate solutions. For $(-,+)$ case, we could not find any plausible local minima.}.
  
 The observed {\it Spitzer} lightcurve is fragmentary and does not cover the caustic-crossing parts. Thus, 
we expect ``color constraints'' might be important to find the correct model including the {\it Spitzer} 
lightcurve \citep[described in Section 5.3 of][]{yee15a}. To incorporate the color constraints, we use 
the ({\it I-H, I}) CMD described in Section 4.4 to find $(I-L)_{18} = 5.157 \pm 0.124$ where the 
subscript $18$ indicates a magnitude system for which $1$ flux unit corresponds to $18$th magnitude. We 
then introduce the ``$\chi^2_{\rm penalty}$'' defined as 
\begin{equation}
\chi^2_{\rm penalty} \equiv{ \left\{ 2.5\log(F_{\rm S,{\it Spitzer}}/F_{\rm S,OGLE})-(I-L)_{18} 
\right\}^{2} \over { \sigma_{\rm cc}^{2} } }
\end{equation}
where $F_{\rm S,{\it Spitzer}}$ and $F_{\rm S,OGLE}$ are the source fluxes of each observatory, 
(i.e. of each passband) conducted from the model. The $\sigma_{\rm cc}$ is the uncertainty of 
the color constraints. The $\chi^2_{\rm penalty}$ increases according to increasing of the 
difference between the model-conducted color and color constraints. Note that, for the 
technical purpose, we additionally increase the penalty defined as 
\begin{equation}
(\chi^2_{\rm penalty})^{\prime}= {\rm fac}^{2} \times (\chi^2_{\rm penalty})
\end{equation}
\begin{equation}
~~if~~ 2.5\log \left( { F_{\rm S,{\it Spitzer}} \over F_{\rm S,OGLE} } \right) > 
\pm{\rm fac} \times \sigma_{\rm cc} (I-L)_{18}
\end{equation}
where ``${\rm fac}$'' is a factor set as $2$. It implies that we use $(\chi^2_{\rm penalty})^{\prime}$ 
that is $4$ times lager than $\chi^2_{\rm penalty}$ if the color difference is lager than $2\sigma$ 
level of the color constraints.  

 In Table \ref{table:two}, we present the best-fit parameters of models both with and without the color 
constraints. The total $\chi^2$ consists of the sum of the $\chi^2$ for the ground-based data and the 
$\chi^2$ for the {\it Spitzer} data; the $\chi^2$ from the color constraint is given separately. For the 
analysis of this event, we find that the color constraints are of minor importance to the model. In fact, 
when fitting without the constraints, the model parameters are recovered within $3\sigma$ of those of the 
best-fit model with the constraints. Especially, the model parameters of the microlens parallax and the 
lens-orbital effect are recovered within $2\sigma$ level. Even though the color constraints play 
a relatively minor role, we conclude the best-fit models of this event are the cases of models considering 
the color constraints (bold parameters in Table \ref{table:two}). Because the color constraint is both an 
intrinsic observable and model-independent, the fact that the best-fit models satisfy these 
constraints serves as an independent check that they correspond to the physical system.

% 4. RESULTS ------------------------------------------------------------------------------------------
\section{Results}

% 4.1. RESULTS: Breaking the degeneracy of the microlens parallax -------------------------------------
\subsection{Breaking the degeneracy of the microlens parallax}

 For the case of the OGLE-2016-BLG-0168 event, we found that two possible solutions $(-,-)$ and $(+,+)$ 
out of the possible four-fold degeneracy for the microlens parallax. The $\Delta\chi^2$ between those 
models is $\sim36.4$, $10.9$ of which comes from $\Delta\chi^2_{\it Spitzer}$. In addition, we found 
inconsistency between the prediction of the lightcurve covered by the {\it Spitzer} data made from the 
ground-based data alone for the $(u_0>0)$ solution and the best-fit model including {\it Spitzer} data 
for the $(+,+)$ solution, as indicated by the different curvatures of the {\it Spitzer} lightcurve seen 
in Figure \ref{fig:two}. Furthermore, for the prediction of the $(+,+)$ case, there exist large 
inconsistencies in the parameters between the $(u_0>0)$ ground-only model and the model including 
{\it Spitzer} data for the $(+,+)$ case at the $4\sigma$ and more than $6\sigma$ levels for the microlens 
parallax and lens-orbital parameters, respectively (see Figure \ref{fig:three}). Hence, considering all 
the clues to resolve the degeneracy, we conclude the $(-,-)$ model is the unique solution that describes 
the nature of the binary lens system of this lensing event.

% 4.2. RESULTS: Confirmation of annual parallax -------------------------------------------------------
\subsection{Confirmation of the annual microlens parallax}

% Figure 4 ----------------------------------------------------------------------------------------
\begin{figure}[ht]
\epsscale{1.10}
\plotone{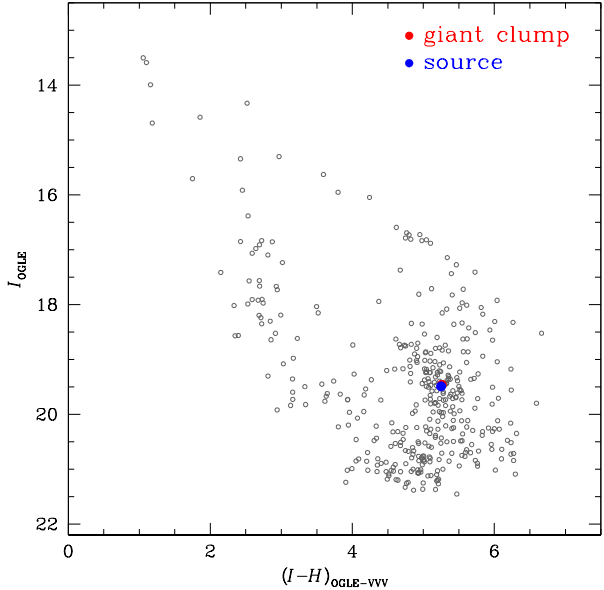}
\caption{\label{fig:four}
(${\it I-H, I}$) Color Magnitude Diagrams of OGLE-2016-BLG-0168 event. The CMD is constructed by 
cross-matching OGLE and VVV data. Red and blue dots indicate the positions of the centroid of giant 
clump and the source star, respectively. 
}\end{figure}
% --------------------------------------------------------------------------------------------------

 As shown in Figure \ref{fig:two}, for the $(-,-)$ case, the prediction is almost the same as the 
lightcurve found by including the {\it Spitzer} data in the fitting. Thus, the higher-order effects 
measured from the ground-based lightcurve alone are confirmed by the {\it Spitzer} observations. 
Note that the prediction of the space-based lightcurve is dominated by the microlens parallax 
parameters. However, the microlens parallax parameters are strongly affected by the lens-orbital 
effect \citep{shin12}. Thus, the lens-orbital parameters are also essential factors for the 
successful prediction of the {\it Spitzer} lightcurve.

 In Figure \ref{fig:three}, we present distributions of the microlens parallax and lens-orbital 
parameters to clearly show the confirmation of the prediction for the ($(-,-)$ and $(u_0<0)$) 
case. We find that parameters of the ground $(u_0<0)$ model that are used for the prediction 
are well matched to those of the model including {\it Spitzer} data for the $(-,-)$ case, i.e. 
within $2\sigma$ and $3 \sigma$ for the microlens parallax and lens-orbital parameters, respectively.

% 4.3. RESULTS: Value of fragmentary Spitzer data -----------------------------------------------------
\subsection{Value of fragmentary {\it Spitzer} observations}

 The confirmation of the microlens parallax and the resolution of the $(-,-)$--$(+,+)$ degeneracy 
show that it is possible to extract valuable information from space-based observations even though 
the observations are fragmentary. Although this is one specific case, it is significant because almost 
all space-based observations have only partial coverage of long time-scale lensing events. Thus, we 
frequently encounter such fragmentary lightcurves.

% 4.4. RESULTS: Properties of the binary lens system --------------------------------------------------
\subsection{Properties of the binary lens system}

 Based on the unique best-fit model, it is then possible to specifically determine the properties of 
the binary lens system. To determine the properties, the angular Einstein ring radius and the microlens 
parallax are essential information. Thanks to good caustic coverage from KMTNet observations, we can 
clearly detect the signal of the finite source effect. From the measurement of $\rho_{\ast}=\theta_
{\ast}/\theta_{\rm E}$, it is possible to determine the angular Einstein ring radius, $\theta_{\rm E}$. 
The angular source radius, $\theta_{\ast}$, can be determined from the position of the source on the 
CMD of the event. The conventional method is to use the ({\it V-I, I}) CMD, but this is impossible in 
this case because the source suffers from severe extinction ($A_I = 4.9$).

 We construct an ({\it I-H, I}) CMD from the OGLE survey and the VISTA Variables and Via Lactea Survey 
\citep[VVV:][]{minniti10} by cross-matching field stars within $60''$ of the source. Based on the CMD 
(see Figure \ref{fig:four}), the centroid of giant clump, which is the reference to measure the 
extinction toward the source, is $({\it I-H, I})_{\rm clump}=(5.26, 19.46)$. The position of the source 
on the CMD is determined as follows. First, from the best-fit model, we have ${\it I}_{\rm S,OGLE}=19.49$. 
Second, based on SMARTS CTIO {\it I-} and {\it H-}band data and converting from $(I_{\rm CTIO}-H_{\rm CTIO})$ 
to $(I_{\rm OGLE}-H_{\rm 2MASS})$ using comparison stars, we find $({\it I-H, I})_{\rm S}=(5.25, 19.49)$. 
As shown in Figure \ref{fig:four}, the locations of the source and the centroid of the red giant clump 
are almost identical. By adopting the intrinsic color and magnitude of the centroid of giant clump 
\citep{bensby13, nataf13} and applying the conventional method \citep{yoo04}, we determine the de-reddened 
color and brightness of the source as $({\it V-I,I})_{\rm 0,S}=(1.05,14.55)$. Finally, the angular source 
radius, $\theta_{\ast}=5.66 \pm 0.40$, is determined by converting $({\it V-I})$ to $({\it V-K})$ based on 
the color-color relation in \citet{bessell88}, and then the angular radius of the source is determined by 
using the color/surface-brightness relation in \citet{kervella04}.

% Table 3 ------------------------------------------------------------------
\begin{deluxetable}{lc}
\tablecaption{The properties of the binary lens system\label{table:three}}
\tablewidth{0pt}
\tablehead{
% ---------------------------------------------------------------------------
\multicolumn{1}{c}{quantity} &
\multicolumn{1}{c}{value} 
}
\startdata
% ---------------------------------------------------------------------------------------------------
Einstein radius, $\theta_{\rm E}$ (mas)              & 1.429 $\pm$ 0.103 \\
Total Mass, $M_{\rm total}$ $(M_{\odot})$            & 0.484 $\pm$ 0.050 \\
Primary Mass, $M_1$ $(M_{\odot})$                    & 0.274 $\pm$ 0.028 \\
Secondary Mass, $M_2$ $(M_{\odot})$                  & 0.210 $\pm$ 0.022 \\
Distance to lens, $D_{\rm L}$ (kpc)                  & 1.572 $\pm$ 0.140 \\
Projected separation, $a_{\perp}$ (AU)               & 2.480 $\pm$ 0.221 \\
Geocentric proper motion, $\mu_{\rm geo}$ (mas/yr)   & 5.573 $\pm$ 0.402 \\
Heliocentric proper motion, $\mu_{\rm hel}$ (mas/yr) & 6.314 $\pm$ 0.456 \\
Stability of system, $ KE/PE^{\dagger}$              & 0.514             \\
% ---------------------------------------------------------------------------------------------------
\enddata
\tablecomments{
$\dagger~$ If the ratio of the Kinetic to Potential energy of the binary lens 
system $(KE/PE)$ is less then $1$, then the orbital motion of binary components 
is physically allowed. However, values $(KE/PE)\sim 1$ and values $(KE/PE)\ll 1$ 
would require very special physical configurations and/or viewing angles. 
Hence, the parameters of the model considering the lens-orbital effect are quite 
reasonable values.
}
\end{deluxetable}
% ---------------------------------------------------------------------------

 Based on the location on the CMD and the intrinsic color of the source, the spectral type of the source an 
early K-type giant. We adopt limb-darkening coefficients based on the classified source type \citep{claret00}. 
The coefficient for {\it I-}band is equal to $\Gamma_{\it I}=(2u/(3-u))=0.5103$ where $u_{\it I}=0.6098$ under 
assumptions of the properties of the early K-type giant: effective temperature, $T_{\rm eff}\sim 4750$ K, 
metallicity, $[M/H]\sim0.0$, turbulence velocity, $V_{\rm t}<2.0~{\rm km/s}$, and surface gravity, 
$\log{g}\sim2.0$. 
 
 Combining the information of the microlens parallax and the angular Einstein ring radius, we can determine 
the properties of the binary lens system according to the equations (1). In Table \ref{table:three}, 
we present the properties of the lens system. The system consists of nearly equal mass stars, 
\begin{equation}
M_1 = 0.27 \pm 0.03~M_{\odot};~M_2= 0.21 \pm 0.02~M_{\odot}, 
\end{equation}
with a projected separation, 
\begin{equation}
a_{\perp} = 2.48\pm0.22~{\rm AU}.
\end{equation} 
The lens system is located $1.57 \pm 0.14$ kpc from us. 

 Since we introduce orbital motion of the lens system, we check whether the best-fit orbital parameters 
are physically reasonable or not. Thus, we derive the ratio of kinetic to potential energy of the system 
to validate the stability of the lens system. The determined value $(KE/PE)\simeq 0.5$ easily satisfies 
the physically bound condition $(KE/PE)<1$. Moreover, it is well away from the regimes $(KE/PE)\sim 1$ 
and $(KE/PE)\ll 1$, both of which would require special geometries and/or viewing angles. Since, 
systematic-induced modeling errors would tend to generate arbitrary values of $(KE/PE)$, the fact that 
the modeling yields a value in the ``typical range'', is further confirmation of its correctness. This 
is important in the present case because lens is unusually close ($D_L=1.6\,$kpc) and the orbital motion 
is usually fast $(|d\alpha/dt|\sim 1\,{\rm radian\,yr^{-1}})$. A number of the most interesting microlensing 
events, e.g. OGLE-2011-BLG-0417 \citep{shin12}, OGLE-2011-BLG-0420 and OGLE-2009-BLG-151 \citep{choi13}, 
are from such nearby lenses, which are intrinsically relatively rare but which frequently permit 
ground-based parallax measurements when they occur. Hence, when one of these can be verified as 
a physically (rather than systematics) generated lightcurve by several independent checks, it enhances 
confidence in this entire interesting class of events.

\begin{figure}[ht]
\epsscale{1.10}
\plotone{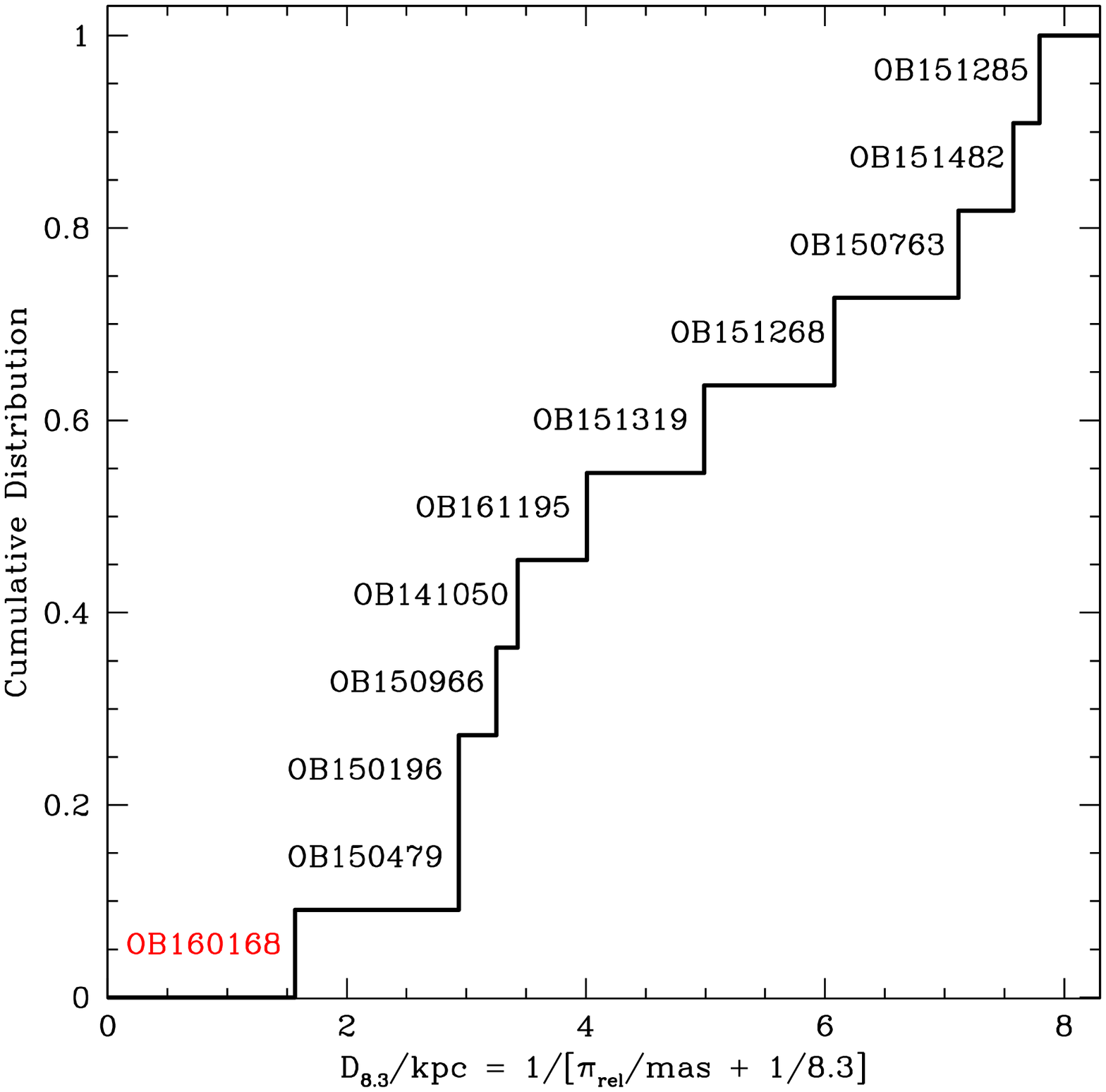}
\caption{\label{fig:five}
Cumulative distribution of $D_{8.3}$ of {\it Spitzer} microlensing events. We adopted $\pi_{\rm rel}$ 
value from published result of each case: OB141050 \citep{zhu15}, OB150196 \citep{han17}, OB150479 
\citep{han16b}, OB150763 \citep{zhu16}, OB150966 \citep{street16}, OB151268 \citep{zhu16}, OB151285 
\citep{shvartzvald15} OB151319 \citep{shvartzvald16}, OB151482 \citep{chung17}, and OB161195 
\citep{shvartzvald17}.
}\end{figure}

In Figure \ref{fig:five}, we present the cumulative distribution of the ``distance parameter'', $D_{8.3}$ 
\citep[see Section 5 of][]{calchi15a}, of published microlensing events with well-measured 
$\pi_{\rm rel}\,(=\pi_{\rm E}\theta_{\rm E}) $ based on {\it Spitzer} observations. We note that the 
lens system of this work is the nearest one with a {\it Spitzer} distance. Assuming that 2-body lenses, 
which dominate this sample, follow the same galactic distribution as all lenses, this distribution 
represents the most precise determination of the {\it Spitzer}-observed lens distance distribution, 
a key factor in understanding the distribution of planets in out galaxy.

%Such nearby 
%lens is believe to be rare. Despite of the rarity, most nearby lenses in the microlensing study tend 
%to show ``special'' characteristics of lens systems determined by ground-based lens-parallax. Thus, 
%the interesting properties of lens systems, i.e., distance to lens and mass which determined by the 
%microlens parallax, might be suspicious due to systematics in ground-based observations. However, the 
%characteristics of this closest binary lens system is quite ``common'', and also the microlens parallax 
%is confirmed by {\it Spitzer} observations. Hence, this work can be an example of nearby lens system to 
%show confirmed case of the lens system.

% 5. SUMMARY & DISCUSSION -----------------------------------------------------------------------------
\section{Summary and Discussion}

 We analyzed the microlensing event OGLE-2016-BLG-0168 based on combined ground- and space-based 
observations obtained from OGLE, KMTNet, and {\it Spitzer} telescopes. It is possible to clearly 
detect signals of higher-order effects in the lightcurve which are caused by the finite source, 
the microlens parallax, and the orbital motion of the binary lens components. Based on the 
additional information from these high-order effects, we found that this event is created 
by a binary system consisting of almost equal mass M-dwarf stars ($\sim 0.27$ and $\sim 0.21$ 
$M_{\odot}$) with a projected separation $\sim 2.5$ AU. The system is located $\sim 1.6$ kpc from us.

 We successfully predict the {\it Spitzer} lightcurve of the $(-,-)$ model case based on the annual 
microlens parallax measured by using the ground-based observations. The annual microlens parallax is 
confirmed at the $2\sigma$ level by the satellite microlens parallax measured with {\it Spitzer} 
observations. In addition, it is possible to resolve the degenerate solutions by using the 
{\it Spitzer} observations. 

% Figure 5 ----------------------------------------------------------------------------------------
% --------------------------------------------------------------------------------------------------

 Our test of the microlens parallax can provide an important example for preparing for the new era of 
microlensing technique in collaboration with space-based observations. In principle, the microlensing 
technique can detect a variety of astronomical objects regardless of their brightness. However, 
additional observables are required to reveal what kind of object produces the microlensing event. 
Among these essential observables, the microlens parallax is one of the key pieces of information that 
reveals the nature of the lens of the event. Thus, it is important to routinely and securely measure 
the microlens parallax. Before the collaboration with space-based observations, measuring the microlens 
parallax usually depended on the time-scale of the lensing event. For some lensing events with long 
time-scale, the microlens parallax signal can be detected. However, this annual microlens parallax 
might be inaccurately measured due to systematics in the data. With the commencement of the era of 
space-based observations collaboration, however, the microlens parallax can be routinely measured 
regardless of the time-scale and magnification level of the lensing event.

 Since most space-based observations cover only part of the full lensing lightcurves with a long 
time-scale due to the relatively short observing window, it is important to conduct a test to 
determine whether it is possible to extract secure information of the microlens parallax or not. 
In addition, since space-based observations can provide a chance to resolve degenerate solutions, 
it is also important to conduct another test to determine whether the degeneracy breaking is 
possible or not by using fragmentary space-based observations. 

We conduct the microlens parallax test by using the fragmentary {\it Spitzer} observation of 
OGLE-2016-BLG-0168 binary lensing event. Our testing result provides an example showing that 
it is possible to verify the microlens parallax and resolve the degeneracy based on space-based 
observations, even though the observation is fragmentary. This result will be helpful for 
preparing collaboration of ground microlens surveys and space telescopes and next-generation 
microlensing survey on the space such as {\it Spitzer} microlensing campaign \citep{yee15b}, 
K2C9 \citep{henderson16}, and WFIRST \citep{spergel15}.

%\mbox{}

\acknowledgments 
This research has made use of the KMTNet system operated by the Korea Astronomy and Space Science 
Institute (KASI) and the data were obtained at three host sites of CTIO in Chile, SAAO in South 
Africa, and SSO in Australia. 
This work is based in part on observations made with the {\it Spitzer} Space Telescope, which is 
operated by the Jet Propulsion Laboratory, California Institute of Technology under a contract 
with NASA.
OGLE project has received funding from the National Science Centre, Poland, grant MAESTRO 
2014/14/A/ST9/00121 to A. Udalski.
Work by A. Gould was supported by JPL grant 1500811.
A. Gould and W. Zhu acknowledges the support from NSF grant AST-1516842.
Work by C. Han was supported by the Creative Research Initiative Program (2009-0081561) of 
National Research Foundation of Korea.
Work by YS and CBH was supported by an appointment to the NASA Postdoctoral Program at 
the Jet Propulsion Laboratory, administered by Universities Space Research Association 
through a contract with NASA. 

\clearpage

\end{document}